\documentstyle[aps,multicol,epsf]{revtex}
\begin{document}
\title{Inelastic Collapse of Three Particles}
\author{Tong Zhou\thanks{Email: tongzhou@control.uchicago.edu}\   and Leo P. Kadanoff\thanks{Email: LeoP@uchicago.edu}\\The James Franck Institute\\The University of Chicago\\Chicago, IL 60637}
\date{\today}
\maketitle
\begin{abstract}
A system of three particles undergoing inelastic collisions in
arbitrary spatial dimensions is studied with the aim of establishing
the domain of ``inelastic collapse''---an infinite number of
collisions which take place in a finite time. Analytic and simulation
results show that for a sufficiently small restitution coefficient,
$0\leq r<7-4\sqrt{3}\approx 0.072$, collapse can occur.  In one
dimension, such a collapse is stable against small perturbations
within this entire range.  In higher dimensions, the collapse can be
stable against small variations of initial conditions, within a
smaller $r$ range, $0\leq r<9-4\sqrt{5}\approx 0.056$.

\vspace{0.15cm}\noindent
PACS: 47.50.+d, 05.20.Dd
\end{abstract}
\begin{multicols}{2}
\section{Introduction}
A system of particles interacting only through inelastic collisions is
a useful idealization of granular materials, and has been much
investigated recently\cite{1,2,3,4,5,6,7,8,9,10}. 
Inelasticity can make such a system
evolve into a collapse state, in which several of the particles
collide an infinite number of times in a finite time interval.

Inelastic collapse in one dimension is well understood\cite{1,2,3,4}. In two
dimensions, McNamara and Young carried out numerical investigations
and found some evidence for inelastic collapses of three particles\cite{5}. To understand the collapse mechanism in higher dimensions, we
study the behavior of three particles in a particular model. Our model
involves collisions which preserve the total momentum and the
components of the momentum perpendicular to the line of centers.  The
component of the relative velocity along the line of centers is
reversed (as in an elastic collision) and reduced by the restitution
coefficient, $0\le r \le 1$.  We look at a situation in which one
particle (labeled zero) takes part in all collisions. The other two
particles (labeled one and two) are alternatively a collider and a
spectator. We assume that all particles have the same mass, and that
particles one and two have identical radii.

There are two possible reasons that previous numerical studies might
have shown collapse. One scenario is that the collapsed state is
represented by one or many attractive fixed points, so that the
collapsing orbit can be stable against small variations in initial
data.  The other scenario is that each orbit is unstable but that the
infinity of collapsing orbits produces an observable collapse
probability. For the specific example of three particles, we find an
attractive fixed point for all dimensions and sufficiently small
coefficient of restitution, $r$. Thus, we establish the possibility of
the first scenario. The second scenario is still possible, but we have
seen no evidence for it.  For a larger $r$, there is an interval in
which the fixed point is unstable against changes in the initial
conditions. 

For dimensions greater than one, after the collapse has occurred,
the particles can separate from one another.  Thus beyond $d=1$,
inelastic collapse is an event, not an end-point, in the
``lives'' of the particles.

To study the collapse, we use the methods of dynamical systems
theory. Specifically, we examine the situation in which particles one
and two are very close to particle zero and aimed so that the
system is very close to the collapsing fixed
point. Figure 1 shows a typical configuration. Particle
$0$ keeps colliding with particle $1$ and particle $2$ repeatedly, and
an inelastic collapse may occur.  After many collisions, the distances
between the particles become small, and the remaining collisions take
place so rapidly that the relative motion of the particles is
small. Therefore, in the inelastic collapse, the angle $\theta$
approaches a limiting value as the number of collision goes to
infinity. We identify the constant-$\theta$ fixed point and then
investigate its stability. We find that a fixed
point of an inelastic collapse exists only when the final $\theta$ obeys:
\begin{equation}
\cos\theta \geq \frac{4\sqrt{r}}{1+r}.
\label{eq:A}
\end{equation}
A collapse state will occur whenever this criterion holds and also
steric effect do not block off the required collisions. (For example
such a blockage will always occur at $\theta = \pi$.) The stability
analysis implies that for the collapse to be stable against small
perturbations in the initial velocities, a stronger condition is required,
namely 
\begin{equation}
\cos\theta >\frac{2\root 3 \of r (1+\root 3 \of r)}{1+r}.
\label{eq:B}
\end{equation}

\section{The collision model}

We use the standard model of inelastic collision: due to a collision
the component of the relative velocity of the colliders along the line
of centers, changes by a factor of $-r$. We denote by $\vec{u}_j$ and
$\vec{x}_j$ the velocity and the position of the $j$th particle at the
instant before a collision occurs. Let us consider a collision between
particles $1$ and $0$.  In the course of the collision, the velocities
of the particles change to:
\begin{eqnarray}
\vec{u}_1' & = & \vec{u}_1 - \vec{\Delta}, \nonumber \\
\vec{u}_0' & = & \vec{u}_0 + \vec{\Delta}, \nonumber \\
\vec{u}_2' & = & \vec{u}_2.
\label{eq:1}
\end{eqnarray}
Here, the momentum transfer is $\vec{\Delta}$. It must point in the
direction of the line of centers. In terms of the coefficient of
restitution, $r$, this transfer is given by the expression
\begin{equation}
\vec{\Delta} = \frac{1+r}{2}  (\vec{x}_1-\vec{x}_0) [(\vec{x}_1-\vec{x}_0)
\cdot
(\vec{u}_1-\vec{u}_0)].
\end{equation}
Here we have assumed that the radii of the colliding particles sum to
unity so that, at the point of collision:
\begin{equation}
(\vec{x}_1-\vec{x}_0)^2 = 1.
\label{eq:3}
\end{equation}

To do dynamical systems theory, we wish to look at the very same
process repeatedly. Therefore, we introduce the superscripts, $c$,
denoting the collider and $s$ denoting the spectator particle, as well as 
a subscript, $n$, to denote the instant before the $n$th collision
occurs (Figure 1). For simplicity, we take the velocity and the position of
particle $0$ to be zero.  In order to make sure that the velocity of
particle $0$ continues to vanish after the collision, we view the
post-collision system from a frame moving with velocity $\vec{u}_0'=
\vec{\Delta}$. Then equations (\ref{eq:1}) - (\ref{eq:3}) read 
\begin{eqnarray}
\vec{u}_{n+1}^s & = & \vec{u}_{n}^c - 2 \vec{\Delta}, \nonumber \\
\vec{u}_{n+1}^c & = & \vec{u}_{n}^s-\vec{\Delta}.
\label{eq:3a}
\end{eqnarray}
These equations are supplemented by the conditions:
\begin{equation}
\vec{\Delta} = \frac{1+r}{2} (\vec{x}_{n}^c) (\vec{x}_{n}^c \cdot
\vec{u}_{n}^c),
\end{equation}
\begin{equation}
(\vec{x}_{n}^c)^2 = 1.
\end{equation}
Additionally, both the velocity and the position of particle $0$
vanish.  Notice that a collider becomes a spectator
immediately following a collision.  The above equations are
complemented by the equations corresponding to the positions of the
particles at the next collision, 
\begin{eqnarray}
\vec{x}_{n+1}^s & = & \vec{x}_{n}^c + t_n \vec{u}_{n+1}^s,  \nonumber \\
\vec{x}_{n+1}^c & = & \vec{x}_{n}^s + t_n \vec{u}_{n+1}^c. 
\label{eq:5}
\end{eqnarray}
The time interval
between the $n$th and the $(n+1)$th collisions, $t_n$, is such 
that the magnitude of $\vec{x}_{n+1}^c$ is unity. 

\section{Flat Surface Approximation}
We now seek fixed points in these equations. We assume that the time
between collisions is sufficiently small so that the $t_n$ terms in equations
(\ref{eq:5}) are negligible and consequently 
\begin{eqnarray}
\vec{x}_{n+1}^s & = & \vec{x}_{n}^c,  \nonumber\\
\vec{x}_{n+1}^c & = & \vec{x}_{n}^s,  
\label{eq:apr}
\end{eqnarray}
during the approach to the fixed point. 

We wish to find a fixed point in the components of the velocity in the
direction of the lines of centers. Specifically, we would like to
investigate how this component decreases in each iteration.
We can define:
\begin{equation}
\vec{x}_{n+1}^c \cdot \vec{u}_{n+1}^c \equiv k_n \vec{x}_{n}^c 
\cdot \vec{u}_{n}^c.
\label{eq:8}
\end{equation}

Taking the dot product of equations (\ref{eq:3a}) respectively into
$\vec{x}_{n}^s$ and $\vec{x}_{n}^c$, and using equations
(\ref{eq:apr}) gives 
\begin{equation}
\vec{x}_{n+1}^s \cdot \vec{u}_{n+1}^s = -r \vec{x}_{n}^c \cdot \vec{u}_{n}^c,
\label{eq:9a}
\end{equation}
\begin{equation}
\vec{x}_{n+2}^c \cdot \vec{u}_{n+2}^c = \vec{x}_{n+1}^s \cdot \vec{u}_{n+1}^s+\frac{1+r}{2}
\vec{x}_{n+1}^c \cdot \vec{u}_{n+1}^c\cos\theta.
\label{eq:9} 
\end{equation}
Equations (\ref{eq:8})-(\ref{eq:9}) thus imply the recursion satisfied
by the scaling factor $k_n$
\begin{equation}
k_{n+1}=-\frac{r}{k_n}+\frac{1+r}{2}\cos\theta.
\label{eq:rok}
\end{equation}
Fixed points can be found by setting $k_{n+1}=k_n$ in (\ref{eq:rok}),
\begin{equation}
k^2 - k \frac{1+r}{2}\cos\theta +r=0.
\end{equation}
As a result, the fixed point of the scaling factor has two
possible values 
\begin{equation}
k_\pm =\frac{1+r}{4}\cos\theta \pm
\sqrt{\left(\frac{1+r}{4}\cos\theta\right)^2-r}.
\label{eq:11}
\end{equation}
Equation (\ref{eq:11}) is one of the major results of our study.

In every collision, the colliding particle must approach particle
$0$. Hence, $\vec{x}_c \cdot \vec{u}_c$ must be negative in 
every iteration. This is possible only if $k$ is a positive real
number. One kind of failure arises when $k$ is complex. Then the real
part of the dot product will change sign infinitely often and no fixed
point can be reached. Thus, for inelastic collapse to occur, the
quantity under the square root in equation (\ref{eq:11}) must be
positive.  This positivity still permits both signs of $\cos\theta$.
However, if the roots are real and the cosine is negative, both roots
will be negative. Hence neither is a possible solution for inelastic
collapse. The only remaining possibility is that inelastic collapse
may occur under the condition on the cosine given by equation
(\ref{eq:A}). According to that statement,  when $r\rightarrow 0$, $\theta$ can
have a value between $0$ and $\frac{\pi}{2}$.  
On the other hand, when $\theta =0$, the well-known
one-dimensional result is recovered \cite{2}, {\it i.e.}, inelastic
collapse is possible for $0\le r< 7-4\sqrt{3}$.  Regions (a) and (b)
in figure 2 are the region of $r$ and $\theta$ for which
we may have inelastic collapse.

We now consider the stability of the above fixed points.  Stability
will imply that a small change in the initial conditions will leave
the system in a collapse state, or in other words, changes will still
permit an infinite number of collisions.  There are two
collapse fixed points distinguished by the values of the two
multipliers $k=k_+$ and $k=k_-$. Subtracting $k_\pm$ from both
sides of (\ref{eq:rok}) yields
\begin{equation}
\frac{k_{n+1}-k_\pm }{k_n-k_\pm }=\frac{k_\mp}{k_n}.
\end{equation}
It is seen that $k_-$ corresponds to an unstable fixed point, while
$k_+$ corresponds to a stable one. Henceforth, we use $k$ to denote
the stable fixed point $k\equiv k_+$.

Next, we investigate the time interval between successive
collisions. Assume that relative motion of the two colliding particles
between each pair of collisions covers a distance which is very small
in comparison to their radii.  Then, we can think of the surface of the
particles as flat. After the $n$th collision, the colliding particle
moves away from the surface and covers a distance $t_n \vec{x}_{n+1}^s
\cdot \vec{u}_{n+1}^s$.  In the next step, this particle moves back
over the same distance, which is given by -$t_{n+1} \vec{x}_{n+2}^c
\cdot \vec{u}_{n+2}^c $ and reaches the surface. Thus, the ratio of
times is
\begin{equation}
\frac{t_{n+1}}{t_n}= \frac{-\vec{x}_{n+1}^s \cdot \vec{u}_{n+1}^s}
{\vec{x}_{n+2}^c \cdot \vec{u}_{n+2}^c}.
\end{equation}
This result may be simplified with the aid of equations
(\ref{eq:8})-(\ref{eq:9a}) to give
\begin{equation}
\frac{t_{n+1}}{t_n}=\frac{r}{k^2}.
\label{eq:14}
\end{equation}
Let $d$ denote the shortest distance between two particles. The
distance ratio equals the product of the 
time ratio and the velocity ratio, $k$, 
\begin{equation}
\frac{d_{n+1}}{d_n}=\frac{r}{k}.
\label{eq:15}
\end{equation}

We performed numerical simulations of the collision process by 
considering three inelastic
particles moving in two-dimension with random initial
conditions. When collapse happens, we compared the ratios
calculated from simulations with the  predictions of  equations
(\ref{eq:8}), (\ref{eq:14}), (\ref{eq:15}).  We found excellent
agreement (Figure 3), indicating that the fixed points
are attractive and indeed correspond to collapse 

\section{Validation of the Flat Surface Approximation}
As observed in the last section, making the approximation
(\ref{eq:apr}) simplifies dramatically the original system and it
can be described by a single ratio $k_n$.  It is as if the particles have
flat surfaces, so that the effect of the tangential components of the velocities of
particles $1$ and $2$ can be ignored. This is true only if the time
intervals are negligible, so that (\ref{eq:apr}) can be
obtained from (\ref{eq:5}).  We will see when such a simplification is
valid, and we will set a criterion for the range of validity of the
approximation.

In our way to equations (\ref{eq:9a})-(\ref{eq:9}), we neglected the
terms like $t_n (u_{n}^c)^2$ in comparison to $\vec{x}_{n}^c \cdot
\vec{u}_{n}^c$, by using the approximation (\ref{eq:apr}).  As noted
above, the former terms decrease as $(\frac{r}{k^2})^n$, while the
latter terms decrease as $k^n$. Thus, the flat surface approximation
is reliable only when $r<k^3$ so that terms proportional to $t_n$
can be safely ignored. This condition can be explicitly written as,
\begin{equation}
r<\left(\frac{1+r}{4}\cos\theta
+\sqrt{\left(\frac{1+r}{4}\cos\theta\right)^2-r}\right)^3,
\end{equation}
which can then be simplified into the form of condition
(\ref{eq:B}). The region of stability determined by this condition is
 region (a) in Figure 2. The maximum possible value of
$r$ for stable behavior is $r_c=9-4\sqrt{5}$.

To this point, our calculation did not rely on the circular
geometry of the particles. The name ``flat surface'' suggests that
when criterion (\ref{eq:B}) is satisfied, 
particles do not experience the curvature of their
surfaces, and collide as if they are flat. This calculation is valid
for arbitrary particle shapes  when (\ref{eq:B}) is satisfied.

We also observe that when criterion (\ref{eq:B}) is satisfied, the
time interval $t_n$ decreases faster than the radial component of the
velocities. In such a situation, collapse happens so quickly that all
other effects, external or internal, have no essential influences to
the process. One can further consider arbitrary interactions
between the particles as well as arbitrary external fields, as
long as all the interactions depends only on the relative positions.
Since the particles' relative positions only change very little during
the process of collapse, all the effects of the interactions on, say,
particle $1$ can be replaced by a constant total force acting on it
which induces an constant acceleration.  This acceleration has very
little effect in the tangential direction since time interval is too
small for it to change the tangential component of the velocity. When the flat
surface approximation is valid, the time interval is even too small
to change the radial velocity component. We conclude that the 
the previously obtained fixed points are unchanged.

\section{Circular Geometry}
In the previous sections, the calculation were performed by neglecting
the $t_n$ terms completely out of equations (\ref{eq:5}) and the fixed
points for inelastic collapse were found when the final state
satisfied condition (\ref{eq:B}). After understanding the
characteristic behavior of the collapse, we can do a more rigorous
calculation to investigate how the system behaves outside the region
satisfying (\ref{eq:B}).

We now see that during the collapse process, the radial velocity components
of particles $1$ and $2$ monotonically decrease  till
they vanish at the moment of singularity, while their tangential
components approach limiting values as the number of collision diverges. 
Hence, we take those tangential components as constants, and
concentrate on the radial components in the calculation of 
fixed points.

For simplicity we study in detail one collision in the situation where
particle $1$ has a zero radius while particle $0$ has a unit radius.
Since the theory depends only upon the sum of the two radii, this case
subsumes all others.  Here we denote the initial instant with a
subscript $i$, and the instant before the collision with $f$. We drop
the superscripts since only particle $1$ is considered. We also assume
the particles are very close, $d\ll 1$, and $-\vec{u}\cdot \vec{x}_i
\ll u_t$, where $u_t$ is the magnitude of the tangential velocity
component of particle $1$.
The collision time is 
\begin{equation}
t=\frac{1}{u_t}\left(\frac{-\vec{u}\cdot \vec{x}_i
}{u_t}-\sqrt{\left(\frac{\vec{u}\cdot
\vec{x}_i}{u_t}\right)^2-2d}\right).
\label{eq:19}
\end{equation}
Immediately before the collision, the radial velocity component of
particle $1$ equals
\begin{equation}
-\vec{u}\cdot \vec{x}_f =\sqrt{(\vec{u}\cdot \vec{x}_i )^2-2du_t^2}.
\label{eq:20}
\end{equation}

Equation (\ref{eq:19}) gives the quantity needed to complete the
equation set (\ref{eq:3a})-(\ref{eq:5}). In order to abbreviate this
calculation, we introduce an effective centrifugal acceleration. In
the above calculation, if we view the situation in a frame rotating
with an angular velocity of $u_t$, around an axis perpendicular to
both $\vec{u}$ and the line of centers, passing through the center of
particle $0$, then particle $1$ has zero tangential velocity, and the
effect of the tangential velocity can be represented by a centrifugal
acceleration $a_1=u_t^2$.  This substitution is justified by noticing
that we can get exactly the same expressions as (\ref{eq:19}) and
(\ref{eq:20}) by using this acceleration. We do not need really use
such a rotating frame. We use $a_1$ as an effective centrifugal acceleration
to replace the tangential component of the velocity with the same
effects. 

After these preparations, we are in a much clearer position. Particle
$1$ and particle $2$ respectively have centrifugal accelerations $a_1$
and $a_2$, which are all in the radial directions. Equation
(\ref{eq:apr}) is again a good approximation. Particle $1$ is moving
on a line, and so is particle $2$. We can further drop the vector
notation. In the following, we use $u$ to denote the radial
component of the velocity of particles $1$ and $2$ immediately 
before a collision, with its positive direction pointing towards the
center of particle $0$.  Consequently, $a_1=-u_{1t}^2$ and
$a_2=-u_{2t}^2$.

The equation set (\ref{eq:3a})-(\ref{eq:5}) reduces to 
\begin{eqnarray}
u_{n+1}^s & = & -ru_n^c+a_{n+1}^st_n \nonumber \\
u_{n+1}^c & = & u_n^s+\frac{1+r}{2}\cos\theta\cdot u_n^c +a_{n+1}^ct_n \nonumber \\
d_n & = & u_{n+1}^c\cdot t_n-\frac{1}{2}a_{n+1}^ct_n^2 \nonumber \\
-d_{n+1} & = & -ru_n^c \cdot t_n+\frac{1}{2}a_{n+1}^st_n^2 \nonumber\\
a_{n+1}^c & = & a_n^s \nonumber \\
a_{n+1}^s & = & a_n^c
\label{eq:set}
\end{eqnarray}
where $d_n$ is the distance between the spectator and particle $0$ at
the instant of the $n$th collision. Recall that $t_n$ is the time
interval between the $n$th and $(n+1)$th collisions.

One simple case can be fully carried through, the case $a_1=a_2\equiv
a$. In this case, particle $1$ and particle $2$ are in a symmetrical
situation so that the recursion relation of the system can be obtained 
after a single collision.

As we did before, we use two nondimensional numbers $k_n$ and
$\alpha_n$ to describe the evolution of the system:
\begin{equation}
k_n\equiv\frac{u_{n+1}^c}{u_n^c}\quad\mbox{and}\quad\alpha_n\equiv\frac{-a\cdot t_n}{u_n^c}.
\end{equation}

Starting from (\ref{eq:set}), after some straightforward calculation,
we get following recursion relation:
\begin{equation}
k_{n+1}=\sqrt{\left(\frac{1+r}{2}\cos\theta
-\frac{r}{k_n}\right)^2-(1+r)\cos\theta \frac{\alpha_n}{k_n}},
\label{eq:27}
\end{equation}
and
\begin{eqnarray}
\alpha_{n+1} & = & \frac{1+r}{2}\cos\theta -\frac{r}{k_n}
 -\frac{\alpha_n}{k_n}\nonumber\\ & - &
 \sqrt{\left(\frac{1+r}{2}\cos\theta
 -\frac{r}{k_n}\right)^2-(1+r)\cos\theta \frac{\alpha_n}{k_n}}.
\label{eq:28}
\end{eqnarray}

Suppose the fixed point is $(\alpha,k)$, then from (\ref{eq:28}), we have
\begin{equation}
\alpha =\frac{-k^2+\frac{1+r}{2}\cos\theta\cdot k-r}{k+1}.
\label{eq:29}
\end{equation}
Since $\alpha\ge 0$ from its definition, the right hand side of the above
equality must be nonegative. Thus, there must exist two real solutions
of $k$ satisfying r.h.s.$=0$. And we readily recover the
condition (\ref{eq:A}).

Substituting (\ref{eq:29}) into (\ref{eq:27}), we have
\begin{eqnarray}
k^4 & = & \left(\frac{1+r}{2}\cos\theta\cdot k-r\right)^2\nonumber\\ &
 & -(1+r)\cos\theta\cdot k\frac{-k^2+\frac{1+r}{2}\cos\theta\cdot
 k-r}{k+1}.
\label{eq:31}
\end{eqnarray}
Of course, $k_\pm$, which appeared before, are solutions of
(\ref{eq:31}). From (\ref{eq:29}) we learn that $(0,k_\pm )$ are fixed
points. Let us look at other solutions of equation (\ref{eq:31}) 
which satisfy
\begin{eqnarray}
k^3+\left(1+\frac{1+r}{2}\cos\theta\right)k^2-\left(r+\frac{1+r}{2}\cos\theta\right)k\nonumber \\
-r=0.
\end{eqnarray}
This equation has one and only one solution of $k$ in the interval
$[0:1]$.  We denote the corresponding fixed point as $(\alpha_0,k_0)$.
Of the three relevant fixed points, $(0,k_-)$ is unstable, and we
concentrate on the stability condition for the other two fixed points.

Suppose that $(\alpha_n,k_n)$ has a small deviation from the fixed point
$(\alpha, k)$. then
\begin{equation}
\left( \begin{array}{c} \delta k_{n+1}\\ \delta\alpha_{n+1}\end{array} \right) =\left( \begin{array}{cc} A_{11} & A_{12}\\A_{21} & A_{22} \end{array} \right) \left( \begin{array}{c} \delta k_n\\ \delta\alpha_n\end{array} \right)
\end{equation}
where
\begin{eqnarray}
A_{11} & = & \frac{1}{2k^3}\left[ 2\left(\frac{1+r}{2}\cos\theta -\frac{r}{k}\right)r+(1+r)\cos\theta\cdot\alpha\right] \nonumber \\
A_{12} & = & -\frac{(1+r)\cos\theta}{2k^2} \nonumber \\
A_{21} & = & \frac{\alpha +r}{k^2}-A_{11} \nonumber \\
A_{22} & = & \frac{(1+r)\cos\theta-2k}{2k^2} \nonumber
\end{eqnarray}
Let us denote by $\lambda$ the eigenvalue of matrix $A$, hence we have
\begin{equation}
\lambda^2-b\lambda+c=0
\label{eq:39}
\end{equation}
where
\begin{eqnarray}
b & = & \frac{1}{2}-\frac{1}{k}+\left[(1+r)\cos\theta-\left(\frac{1+r}{2}\cos\theta\right)^2\right]\frac{1}{2k^2}\nonumber\\
 & & +r(1+r)\cos\theta\cdot\frac{1}{k^3}-\frac{3r^2}{2k^4},\nonumber\\
c & = & \frac{r^2}{k^5}.\nonumber
\end{eqnarray}
For the fixed point $(0,k_+)$,
\begin{equation}
b=\frac{r}{k_+^3}(1+k_+)\quad\mbox{and}\quad c=\frac{r^2}{k_+^5}.
\end{equation}
So
\begin{equation}
\lambda_1=\frac{r}{k_+^2}\quad\mbox{and}\quad \lambda_2=\frac{r}{k_+^3}.
\label{eq:eig}
\end{equation}
The point $(0,k_+)$ is an attractive fixed point if and only if
$r<k_+^3$.  Hence we recover the condition of (\ref{eq:B}). If we only
require $\lambda_1<1$, we can recover condition (\ref{eq:A}).

For the fixed point $(\alpha_0,k_0)$, we find 
that $\alpha_0>0$ if and only if the condition of (\ref{eq:B}) is satisfied, but
that condition also decides the range of $(r,\theta)$ inside which we
will have
\begin{equation}
1-b+c<0
\end{equation}
From equation (\ref{eq:39}) we know that one eigenvalue of matrix $A$ is
larger than $1$, implying that the corresponding fixed point is
unstable.  We conclude that there are no additional stable fixed
points beyond those which satisfy condition (\ref{eq:B}), and there
are no stable collapses outside that range in the case $a_1=a_2$.

We believe this is true for the general situation of $a_1\ne a_2$.
When condition (\ref{eq:B}) is violated, our simulations show that
even though the particles can be very close, they will get apart
before having collided an infinite number of times.  

From the above
calculation, and specifically equation (\ref{eq:eig}), we can see the
parameter space $(r,\theta)$ can be divided into three regions
(Figure 2):

a) When condition (\ref{eq:B}) is satisfied, both the eigenvalues of
matrix $A$ are smaller than unity. The fixed point is stable in all
directions in space $(\alpha, k)$. It is the collapse region.

b) When condition (\ref{eq:B}) is violated but the condition (\ref{eq:A})
is satisfied, the eigenvalue $\lambda_2>1$, while $\lambda_1<1$. In
this region, particles can have any number of collisions before they
might eventually separate.

c) When the condition (\ref{eq:A}) is violated, both eigenvalues
of matrix $A$ are larger than unity. Collapse does not occur.

The above calculation is independent of the sources of the
accelerations, which could be the interactions between particles. The
calculation confirms our previous argument that interactions are
irrelevant in the process of collapse\footnote{But when there is a
sufficiently strong attraction, e.\,g.\ gravity, between particles so
that the directions of the accelerations are reversed, we want fixed
points with $\alpha\le 0$. Then the fixed point $(\alpha_0,k_0)$
becomes stable when condition (\ref{eq:B}) is violated. Inelastic
collapse can happen in a much larger region of $(r,\theta)$.}.

We also observe that in the above calculation, the circular geometry of
the particles is not essential. The calculation is also valid for
arbitrary shape of the particles, with the corresponding radius of
curvature replacing the radii used. Even though the radius of
curvature is not relevant in deciding the behavior of collapse
process---the radii do not show up in the expression of the stable
fixed point or condition (\ref{eq:B}), it does have some effects. The
centrifugal acceleration, which obviously is important in deciding the
probability of collapse, is inversely proportional to the radius. So
chances are larger for collapse to happen when the colliding point is
at a position on the surface with a larger radius of curvature.

\section{Conclusion}
We demonstrated analytically the existence of inelastic collapse for
three particles in all dimensions. At the last moment of collapse, the
three particles have a cyclic behavior, which is characterized by a
fixed point. We have established the range of the parameters for which
the fixed point exists and the range for which it is stable.

\subsection*{Acknowledgments}
We would like to express our thanks to Y.\,Du, S.\,Esipov,
H.\,M.\,Jaeger, M.\,Mungan, S.\,R.\,Nagel, N.\,Sch\"orghofer and
W.\,R.\,Young for very helpful discussions and especially to Eli
Ben-Naim for continued advice and discussions. This work was supported
in part by NSF-DMR and in part by DOE.

\end{multicols}

\begin{thebibliography}{99}
\bibitem{1} B. Bernu and R. Mazighi, {\sl J. Phys.\ A} {\bf 23}, 
5745 (1990).

\bibitem{2} S. McNamara and W. R. Young, {\sl Phy.\ Fluids A} {\bf 4},
496 (1992).

\bibitem{3}E. Clement, S. Luding, A. Blumen, J. Rajchenbach and 
J. Duran, {\sl Int.\ J. Mod.\ Phys.\ B} {\bf 7}, 1807 (1993).

\bibitem{4}P. Constantin, E. Grossman and M. Mungan, 
{\sl Physica D} {\bf 83}, 409 (1995).

\bibitem{5} S. McNamara and W. R. Young, {\sl Phy.\ Rew.\ E} {\bf 50},
R28 (1994).

\bibitem{6}P. K. Haff, {\sl J. Fluid Mech.} {\bf 134}, 401 (1983).

\bibitem{7}M. A. Hopkins and M. Y. Louge, {\sl Phy.\ Fluids A} 
{\bf 3}, 4 (1990).

\bibitem{8}I. Goldhirsch and G. Zanetti, {\sl Phys.\ Rev.\ Lett.}
{\bf 70}, 1619 (1993).

\bibitem{9}Y. Du, H. Li and L. P. Kadanoff, {\sl Phys.\ Rev.\ Lett.}
{\bf 74}, 1268 (1995).

\bibitem{10}H. M. Jaeger, S. R. Nagel and R. P. Behringer, ``The
Physics of Granular Materials'', preprint.
\end{thebibliography}
\end{document}